\documentstyle[aps]{revtex}

\input epsf

\begin{document}
\title{Percolative Shunting on Electrified Surface}
\author{Yu. I. Kuzmin}
\address{Ioffe Physical Technical Institute of Russian Academy of Sciences,\\
Polytechnicheskaya 26 St., Saint Petersburg 194021 Russia\\
e-mail: yurk@shuv.ioffe.rssi.ru; iourk@usa.net\\
tel.: +7 812 2479902; fax: +7 812 2471017}
\date{\today}
\maketitle

\begin{abstract}
The surface discharge of electrified dielectrics at high humidity is
considered. The percolative nature of charge transport in electrets is
established. Particular attention is given to the phenomena of adsorption
and nucleation of electrically conducting phase in the cause of percolation
cluster growth on electrified surface. The critical index of the correlation
lenght for percolation cluster is found, and its value is in good agreement
with the known theoretical estimations.
\end{abstract}

\section{Introduction}

The percolation theory is most useful in dealing with charge transport when
the local electric conductivity has a large dispersion. Such a situation is
inherent in surface discharge of electret under the action of humidity of
the ambient atmosphere. It is just the surface discharge which is
responsible for electret stability under the real operating conditions for
electronic devices, whereas the charge transport through a bulk of
dielectric occurs mainly when physical properties of electret materials are
studied, especially, using various techniques of thermally stimulated
discharge .

The basis for consideration is formed by the percolation model of electret
charge relaxation originally proposed in \cite{k1}. This model is based on
the combined application of percolation theory \cite{b2}-\cite{s4} and the
Kolmogorov model \cite{k5} for 2D nucleation kinetics. Such an approach has
allowed to get a quantitative description for charge decay on the
electrified surface under the action of electrically conducting external
agent, primarily, the ambient air humidity.

\section{Basic Concepts of Percolation Approach to Charge Relaxation in
Electrets}

The water adsorption gives rise to the conducting channels on the
electrified surface that modifies significantly its electric properties.
Since water is much more better electrical conductor than the electret
material, these channels shunt the surface of the specimen. Such a physical
situation cannot be described correctly in terms of the usual specific
surface resistance of the electret material, because the uniformity of
surface conductivity is perturbed strongly. When the surface fraction
covered by the adsorbate reaches the percolation threshold, the conducting
channels will form an infinite percolation cluster along which the excess
charge flow off to be subsequently neutralized. At the instant an infinite
cluster is arisen, the percolation transition occurs. This cluster shunts
the entire surface of the specimen, so all the parts of electrified surface
that are connected to it does not contribute to the surface potential of the
electret $U$. Thus we can write: $U\propto 1-P$, where $P$ is the
percolation cluster density. The latter quantity fulfills the role of an
order parameter in percolation problems, and gives the probability that a
randomly located point of the surface belongs to an infinite cluster. The
percolation cluster density depends on the fraction of the surface occupied
by adsorbate $\theta $ at time $t$: $P=P\left( \theta \right) $, where $%
\theta =\theta \left( t\right) $. To find the adsorption kinetics we shall
use the Kolmogorov theory \cite{k5}, which has been developed for a
treatment of the nucleus growth in steady-state reaction space of an
arbitrary dimensionality. All initial assumptions of this theory are assumed
to be satisfied for electret discharge process. For the case of 2D nucleus
growth this means that: (i) the reaction space is unbounded, so that the
mean area of incipient nuclei is infinitely small compared to the total area
of electret surface; (ii) Poisson law holds for the nucleation process, so
all the nucleation centers are generated uniformly on the electret surface
in a random way at a finite rate of $\alpha =\alpha \left( t\right) $ for
the expected number of arising centers per unit time and per unit area;
(iii) all of the nuclei have the same convex shape so they are geometrically
similar; (iv) the growth is uniform so at any stage of the growth process
all the nuclei have the same rate of the growth, which is a function of
time: $\upsilon =\upsilon \left( t\right) .$

Under these assumptions we have the following kinetics for the growth of
adsorbed phase for all time after electrization occurred at time $t=\tau
_{0} $: 
\begin{equation}
\theta \left( t\right) =1-Q\left( \tau _{0},t\right) \exp \left(
-\int\limits_{\tau _{0}}^{t}d\zeta \alpha \left( \zeta \right) S\left(
R_{2}\left( \zeta ,t\right) \right) \right)  \label{Eq1}
\end{equation}
where 
\begin{equation}
Q\left( \tau _{0},t\right) \equiv \exp \left( -\int\limits_{0}^{\tau
_{0}}d\zeta \alpha \left( \zeta \right) S\left( R_{1}\left( \zeta ,\tau
_{0}\right) +R_{2}(\tau _{0},t)\right) \right)  \label{Eq2}
\end{equation}

\bigskip

The functional $R(\zeta ,t)\equiv \int\nolimits_{\zeta }^{t}\upsilon (\eta
)d\eta $ gives the radius at the time $t$ of the nucleus arisen at time $%
\zeta $; $S=S\left( R\right) $ is the area of an isolated nucleus of radius $%
R$; $\alpha =\alpha \left( t\right) $ is the nucleation rate. The
relationship $R=R\left( \zeta ,t\right) $ specifies the geometrical features
of adsorbate nucleus growth. The nucleus arisen after electrization has the
radius $R=R_{2}\left( \zeta ,t\right) $; otherwise, the radius is $%
R=R_{1}\left( \zeta ,t\right) $ while the nucleus grows before
electrization, and $R=R_{1}\left( \zeta ,\tau _{0}\right) +R\left( \tau
_{0},t\right) $ when its growth is going on thereafter. The function $%
Q\left( \tau _{0},t\right) $, as given by equation (\ref{Eq2}), describes
the growth of all the nuclei arisen before electrization which continue to
grow at a later time. The value $Q\left( \tau _{0},t\right) $ is equal to
the fraction of electret surface remained clear of such nuclei at time $t$.
Over a period of time from the preparation of the future electret material ($%
t=0$) until the electrization ($t=\tau _{0}$) the interaction with the
ambient medium is taking place. After thermodynamic equilibrium has been
established, the fraction of the surface occupied by the adsorbate will
remain constant provided that the specimen is kept in controlled
steady-state conditions. In this case $Q\left( \tau _{0},t\right) $ may be
replaced by the time-independent function $Q\left( \tau _{0}\right) $, which
gives the fraction of the surface left free of adsorbate before the
electrization. A large amount of new nucleation centers are generated in the
course of electrization, so it is precisely these ones that greatly
reinforce adsorption process, which is activated by the proper electric
field of electret itself. A good agreement with the experimental data on the
surface potential decay of teflon electrets has been achieved using such a
simple representation for the nucleation rate: $a\left( t\right) =\beta
\delta \left( t-\tau _{0}\right) +\alpha $, where $\beta $ is the
concentration of nucleation centers appeared in the course of electrization, 
$\delta \left( t\right) $ is Dirac pulse function, $\alpha _{0}$ is the
constant rate of spontaneous nucleation, which is going on over the entire
process of electret discharge. Thus, both instantaneous nucleation during
electrization and the continuous one during subsequent storage (or, really,
operation) of the electret are taken into account.

Thus, upon shifting in time by the interval $\tau _{0}$, the Equation (\ref
{Eq1}) may be simplified as 
\begin{equation}
\theta \left( t\right) =1-Q\left( 0\right) \exp \left(
-\int\limits_{0}^{t}d\zeta \alpha \left( \zeta \right) S\left( R_{2}\left(
\zeta ,t\right) \right) \right)  \label{Eq3}
\end{equation}

Now electrization is assumed to be over at time $t=0$ so the discharge is
started just thereafter. When the electret surface fills with adsorbate, a
percolation cluster forms there. As soon as an infinite cluster appears, a
percolation transition takes place so the surface potential of electret
starts to decay. There are three different types of percolation transitions.
The first one corresponds to the case when infinite cluster forms after a
lapse of some time since electrization. Up to this moment the fraction of
the surface covered by adsorbate is below the percolation threshold, so the
surface potential of the electret remains constant and the greatest
stability of the electret charge is thereby achieved. For percolation
transition of the second type the infinite cluster arises just by the end of
electrization, therefore the surface potential begin to decay at once, with
no plateau on its time-dependence curve. In the case of percolation
transition of the third type the infinite cluster has already existed during
electrization, resulting in a catastrophic discharge of electret
(percolation breakdown). The percolation transitions of all the
above-mentioned kinds were observed experimentally \cite{c6}.

\section{Scaling of Surface Potential near Percolation Threshold}

The distinctive feature of percolation cluster growth is that adsorption of
conducting phase is activated by electric field: the polar molecules of
water are attracted by the additional adsorption centers arising from the
charges on electrified surface. Immediately after electrization dielectric
is in highly non-equilibrium state, so the relaxation processes are going
especially fast. As the electrified surface is getting covered by the
network of conducting channels, the further discharge slows down since the
appearance of new paths for current flow has been hampered. The curve of
electret discharge has a saturation then: the surface potential of electret
first falls off steeply, whereupon it stays on some stable level for a long
time. Such a saturation of the surface potential occurs when the correlation
length of percolation cluster becomes much shorter than the characteristic
geometrical size of the electret, so that the backbone network of infinite
cluster has sufficiently small cells.

The percolation nature of the surface discharge of electrets is confirmed by
the good agreement between the experimental and theoretical values of the
critical index of the correlation length. According to the scaling
conception \cite{s4}, \cite{g7}-\cite{s9} the correlation length is the only
geometrical size that is intrinsic to percolation cluster near the
percolation
\begin{figure}[tbp]
\epsfbox{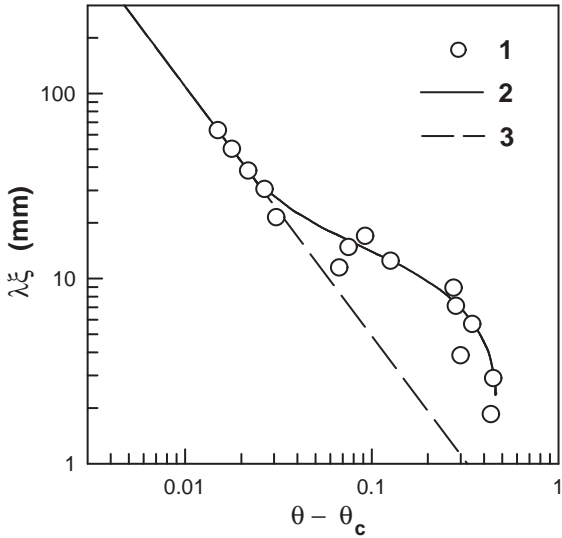}
\caption{Estimation of the correlation length critical index. (Points (1)-
experimental data, curve (2) - optimal approximation, dotted line (3) -
scaling asymptote).}
\label{figure1}
\end{figure}
threshold. Let us suppose that the correlation length $\xi $ at
the saturation of surface potential is proportional to the characteristic
size of electret $D$. This allows us to define the surface potential $U_{s}$
at the saturation point as $U_{s}=U\left( t=t_{s}\left| \xi \propto D\right.
\right) $, where $t_{s}$ is the time when saturation is reached. Thus, the
correlation length at the saturation point is equal, up to a constant factor 
$\lambda $, to electret size: $D=\lambda \xi $. Hence it follows that the
larger is electret, the greater is its surface potential at saturation.
Furthermore, the surface potential is uniquely determined by the percolation
cluster density, which is given by the fraction of electret surface covered
by adsorbate. So it is possible to find the relationship between $\lambda
\xi $ and $\theta $ from experimental data for the discharge of electrets of
various sizes. Such data have been obtained for teflon electrets exposed to
98\%-humidity at room temperature over a period of three months. Teflon has
been chosen as one of the best electret material. The results are shown in
Fig.~\ref{figure1}. 

All the points have been plotted after abscissa conversion from the surface
potential at saturation $U_{s}$ to the difference between the fraction of
the surface covered by adsorbate $\theta $ and the percolation threshold $%
\theta _{c}$ for this quantity. ($\theta _{c}=0.5$ for the case of 2D
continual percolation). The initial surface potential after electrization
was equal to 300V for all the specimens, whereas its value at saturation has
fallen within the range between 30V and 210V according to the electret size.

The correlation length near the percolation threshold obeys the following
scaling law \cite{g7}, \cite{k8}: 
\begin{equation}
\xi \propto \left| \theta -\theta _{c}\right| ^{-\nu }\text{ \ \ \ \ \ \ \ \
\ \ \ , \ \ \ \ \ \ \ \ \ \ \ \ }0<\theta -\theta _{c}<<1  \label{Eq4}
\end{equation}
where $\nu $ is the critical index of the correlation length for 2D
percolation. Thus, the value of this index can be found from the slope of
the scaling curve $\lambda \xi $ vs $\left( \theta -\theta _{c}\right) $
plotted in logarithmic scale. An appropriate procedure is illustrated in
Fig.~\ref{figure1}. The optimal approximation allows us to get the slope of
the scaling asymptote in the region near a percolation threshold, where
scaling behavior of Eq.~(\ref{Eq4}) is valid. The critical index of the
correlation length estimated in this way was $\nu =1.4\pm 0.1$, which is in
good agreement with the known theoretical estimates for 2D percolation \cite
{s4}, \cite{g7}, \cite{k10}.

Scaling behavior of the surface potential near a percolation threshold is
closely related to the peculiarities of adsorbate nucleus growth. In the
present work the charge transport on electrified surface has been studied
for the different laws of nucleation. The relationship between the local
growth rate of conducting phase nuclei and the integral propagation velocity
of the potential jump attendant on the formation of conducting channels on
the electret surface has been cleared up. This potential jump travels as
fast as the different points of electrified surface are getting electrically
connected when adsorption is going on. It has been found that the integral
velocity of potential jump propagation $V$ far exceeds the local growth rate
of an isolated nucleus $\upsilon $: $V=\left( 2n-1\right) \upsilon $, where $%
n$ is the number of nucleation centers over a distance $L$ traveled by the
potential jump, which can be written as $L(t)=\int\nolimits_{0}^{t}V(\eta
)d\eta $. Further, the utmost distance that the potential jump can reach on
the electrified teflon surface at high humidity has been measured. The data
for the water adsorption on teflon have been analyzed by means of three
principal adsorption isotherms: of Brunauer-Emmet-Teller, Langmuir, and
Henry. It has been found that the growth rate of adsorbate nucleus is well
approximated by the exponential law: $\upsilon \left( t\right) =\upsilon
_{0}\exp \left( -t/\tau \right) $, where $\tau $ is a relaxation time, and $%
\upsilon _{0}\equiv 1/\left( 2\tau \beta ^{1/2}\right) $. For exponential
relaxation of the growth rate the utmost distance $L\left( \infty \right) $
traveled by the potential jump over an infinitely long period of time is
finite; whereas if the growth rate was constant or obeyed the hyperbolic
decay, that propagation length would be infinitely large. Exponential
relaxation of the growth rate has been also verified directly by the fact
that experimental data for the utmost distances of potential jump
propagation can be linearized in the co-ordinates: $\left\{ \ln \left(
L\left( \infty \right) -L\left( t\right) \right) \text{ vs }t\right\} $.
Thus, the found exponential law for the growth rate adequately describes the
main features of the charge transport on electrified teflon surface.

\section{Conclusion}

The charge transport on an electrified surface is of a percolative nature as
a result of the adsorption of a conducting phase stimulated by the electric
field. The percolation approach has allowed to get a quantitative
description of the surface discharge of electrets, so it is possible now to
predict the stability of electret devices under the real operating
conditions.

\end{document}